\documentstyle[aaspp4,12pt,psfig]{article}
\lefthead{Bower \& Backer}
\righthead{7 mm $\lambda$ VLBA Observations of Sgr A*}
\begin{document}

\newcommand\degd{\ifmmode^{\circ}\!\!\!.\,\else$^{\circ}\!\!\!.\,$\fi}
\newcommand{\etal}{{\it et al.\ }}

\title{7 mm $\lambda$ VLBA Observations of Sagittarius A*}
\author{Geoffrey C. Bower}
\affil{Max Planck Institut f\"{u}r Radioastronomie, Auf dem H\"{u}gel 69, D 53121 Bonn Germany}
\author{Donald C. Backer}
\affil{Astronomy Department \& Radio Astronomy Laboratory, University of California, Berkeley, CA 94720}

\begin{abstract}

We present 7 mm $\lambda$ VLBA observations of the compact nonthermal radio
source in the Galactic Center, Sgr A*.  These observations confirm the
hypothesis that the image of Sgr A* is a resolved elliptical Gaussian caused by
the scattering of an intervening thermal
plasma.  The measured major axis of Sgr A* is $0.76 \pm 0.04$ mas, consistent
with the predicted scattering size of $0.67 \pm 0.03$.  We find an axial
ratio of $0.73 \pm 0.10$ and a position angle of $77\degd 0 \pm 7\degd 4$.
These results are fully consistent with VLBI observations at longer
wavelengths and at 3 mm $\lambda$.
We find no evidence for any additional compact structure to a limit of
35 mJy.   

The underlying radio source must be smaller than 4.1 AU for a galactocentric
distance of 8.5 kpc.  This result is consistent with the conclusion that
the radio emission from Sgr A* results from 
synchrotron or cyclo-synchrotron radiation
of gas in the vicinity of a black hole with a mass near $10^6 M_{\sun}$.

\end{abstract}

\keywords{Galaxy: center --- galaxies: active --- scattering }

\section{Introduction}

The radiation from 
the compact radio source in the Galactic Center, Sgr A*,
has been clearly determined to be
nonthermal in nature.  VLBI observations have found
the source to have a brightness temperature in excess of
$1.4 \times 10^{10}$ K  and a size less than 1 AU (Rogers \etal 
\markcite{roger94} 1994).  
The compact source has been interpreted as a  black hole
undergoing accretion at a very low rate 
($\sim 10^{21}$ to $10^{22} {\rm\ g\ s^{-1}}$), either through
advection dominated accretion (Narayan, Yi \& Mahadevan \markcite{naray94}
1994) or through 
spherical accretion which may include a disk at small radii (Melia 
\markcite{melia94} 1994).  
In both models the accreting material is
mass lost from nearby massive stars through
winds.  The radio emission may originate in 
an infalling spheroid, a disk or
a low-power Blandford-K\"{o}nigl jet
(Falcke, Mannheim \& Biermann \markcite{falck93} 1993).
These theories are constrained by the
broadband spectrum and the compact size of the radio source.

Recent dynamical evidence supports the black hole hypothesis.
The proper motions of early type stars 
within 0.01 pc of Sgr A* are greater than 1000 km s$^{-1}$ (Genzel
\etal \markcite{genze97} 
1997).  The velocity dispersion as a function of radius
suggests a contained mass of $2.6 \pm 0.35 \times 10^6  M_{\sun}$.
Further, if the kinetic energies of these stars are in equipartition
with that of Sgr A*, then we find that
the proper motion of Sgr A* (Backer \markcite{backe96} 1996)
implies a minimum mass of  $1.4 \times 10^4 M_{\sun}$.  
If this mass is confined
within 1 AU as the VLBI results imply, then a mass density of
$3 \times 10^{19} {\rm M_{\sun} pc^{-3} }$
is required.  Considerations of the short time scales of stability
of a cluster of dark objects with this density lead to the strong
conclusion that Sgr A* indeed harbors a massive black hole
(Maoz \markcite{maoz97} 1997).

Due to diffractive scattering by electrons in the turbulent
interstellar medium near the Galactic Center, the angular
size of Sgr A* is broadened (e.g., Yusef-Zadeh \etal \markcite{yusef94}
1994 and
references therein).  
The source is elliptical at a position angle of approximately $80\arcdeg$ 
with an
axial ratio near 0.5.  The major axis has a $\lambda^{2.01 \pm 0.02}$
dependence that extends from 20 to 0.3 cm.
The scattering hypothesis is supported by the discovery of a similar
angular broadening and asymmetry in the images of OH masers within 
15 arcmin of Sgr A* (van Langevelde \& Diamond \markcite{vanla91} 1991).
Since the effect of scattering decreases more rapidly with 
wavelength than angular resolution, at a short enough wavelength
the compact source should appear unobscured.

Currently VLBI observations at millimeter wavelengths stand in conflict.
At 3 mm $\lambda$ Rogers \etal \markcite{roger94} (1994)
found an upper limit to the apparent size of 0.2 mas,
in agreement with an extrapolation of the scattering law.  Observing
at 7 mm $\lambda$ in August 1992,
Backer \etal \markcite{backe93} (1993) found with 5 stations of the 
National Radio Astronomy Observatory\footnote{
The National Radio Astronomy Observatory
is a facility of the National Science Foundation operated under
cooperative agreement by Associated Universities, Inc.}
Very Long Baseline Array
(VLBA) 
an apparent size of 0.7 mas, also in agreement
with the scattering law.  However, Krichbaum \etal \markcite{krich93}
(1993), observing with an array of 4 VLBA 
stations at 7 mm $\lambda$ in May 1992, 
found a size $1.7 \times 0.7$ mas at a position angle of $-20 \arcdeg$.
They also consider a two component model  in which the brighter
component has a size of $ 0.7 \pm 0.1$ mas and the fainter component
is at a position angle of $-25\arcdeg$.  

In this paper we present 7 mm $\lambda$ observations of Sgr A* using
the full VLBA.  The additional baselines in these observations
significantly improve the density and extent of $uv$ coverage
over the previous experiments.
We present a map as well as a study of the visibility data.

\section{Observations and Data Reduction}

The VLBA
observed Sgr A* on 1994 September 29 at 43 GHz with 
a bandwidth of 64 MHz.  The data were correlated 
in Socorro, New Mexico.
The compact blazar NRAO~530 was used as an amplitude calibrator and
fringe detection source.  
High SNR detections of NRAO~530
were made on all baselines.  

Initial
reduction of the data was performed with the Astronomical Image
Processing System (AIPS).
{\it A priori} amplitude calibration was performed first.
We determined that the zenith atmospheric optical depth ranged from 0.05
at Mauna Kea to 0.18 at Hancock.
Singleband delays were determined with fringe fitting
to a short segment of NRAO~530 data and then applied to 
the Sgr A* visibilities.  Multiband delays and rates were
found by fringe fitting directly to the Sgr A* visibilities.
The SNR of detections and the
consistency of fringe rate and multiband delay solutions
indicate that detections were made to all stations except
Mauna Kea, Saint Croix and Hancock.  No detection of Sgr A* was made at a
$uv$ distance greater than 250 $M\lambda$.
Data were averaged to 64 seconds after fringe detection.

We improved the amplitude calibration of Sgr A* through mapping
of NRAO~530.  A single Gaussian with a size of 0.102 by 0.056 mas 
fits the NRAO~530 visibilities very well.  
Station gains deviated from unity by 20\%
or less.  These station gains were applied to the visibility
data of Sgr A* and used for imaging and visibility analysis.
This calibration significantly reduced the scatter in the Sgr A*
visibility amplitudes.

We self-calibrated the visibility phase
of Sgr A* with iterative mapping.
We show in Figure~\ref{fig:sgra} an image of Sgr A* made with
a robustness parameter of 2
(Briggs \markcite{brigg95} 1995).  The corresponding beam FWHM 
is $2.230 \times 0.720$ mas at a position angle of $12\degd 9$.
We show the same image with the CLEAN components
convolved with a circular beam with FWHM of 0.400 mas in
Figure~\ref{fig:sgrasup}.

\section{Results}

\subsection{Size and Flux of Sgr A*}

We fit Gaussian models directly to the calibrated visibility data with
a non-linear least-squares method.
We only used
visibility data on baselines shorter than 200 $M\lambda$.
We show in Figure~\ref{fig:sgrauvd}
the visibility amplitudes as a function of $uv$ distance along 
with circular Gaussian models.  
A noise bias of 0.140 Jy was added in quadrature to the model.
The best circular Gaussian fit has a 
FWHM of $0.727 \pm 0.033$ mas.  
An elliptical Gaussian fit
gives a FWHM of $0.762 \pm 0.038$ mas in a position angle of
$77\degd 0 \pm 7\degd 4$.  We find an axial ratio of $0.73 \pm 0.10$.
We find consistent results fitting Gaussians to the two
images.

The absence of a detection on the Hancock-Saint Croix baseline
provides an opportunity to place an upper limit on the axial
ratio.  This baseline has moderate North-South resolution with
$v=200 M\lambda$ at the beginning of the track.  The average amplitude
on this baseline coherently integrated over the fringe detection
time scale is $\sim 140$ mJy.
We find the axial ratio $a \lesssim 1.0$.

The total flux determined from the fit to the visibility data is 
$1.28 \pm 0.10$ Jy.  The quoted error does not include systematic
calibration errors, which may be as high as 20\%.

\subsection{Is There Asymmetric Structure?}

In order to search further 
for non-elliptically symmetric structure in Sgr A*, we
analyzed the closure phases in detail.  We extracted the visibility data
from AIPS and formed the closure phase in 2 second intervals.  We vector
averaged this complex phasor for
10 minute segments.  Errors were determined from the distribution
of the closure phases (Thompson, Moran \& Swenson \markcite{thomp91}
1991, section 9.3).  
For a weak signal, the distribution is
linear in the cosine of the phase and the phase error goes as
\begin{equation}
\sigma_\phi= {\pi \over \sqrt{3} } \left( 1 - \sqrt{ 9 \over 2 \pi^3}
{V \over \sigma} \right),
\end{equation}
where $V$ is the bispectrum amplitude and $\sigma$ is the noise amplitude.
For a strong signal, the distribution is Gaussian and we find
\begin{equation}
\sigma_\phi={\sigma \over V}.
\end{equation}

We computed the $\chi^2$ statistic for the Krichbaum \etal \markcite{krich93}
(1993)
two component model (their model B) and for a symmetric model.  We find
$\chi^2_{Krichbaum}=2104$ and $\chi^2_{symmetric}=449$
for $N=385$ degrees of freedom, which clearly eliminates the
Krichbaum \etal model at the epoch of these observations.  

We show some of the closure phases along with
the Krichbaum \etal model in
Figure~\ref{fig:sgracp1}.
The closure phase plots reveal that the phase does not deviate 
significantly from zero where the Krichbaum \etal model predicts
substantial variation.  These deviations are apparent on both
the close inner triangles and on the more spread out triangles
which include BR, FD and OV, stations not used in the Krichbaum \etal
experiment.

Is there any evidence for asymmetry aside from that in
the Krichbaum \etal model?  The reduced $\chi^2$ value 
for the symmetric hypothesis is 1.17 which leaves little phase
space for a secondary component.   
The images and the visibility data 
do not reveal any evidence for asymmetric structure at
the level of the beam sidelobes, 35 mJy.
The diffuse components in Figures~\ref{fig:sgra} and \ref{fig:sgrasup}
coincide with
beam sidelobes and are, therefore, unlikely to be real.

\section{Conclusions}

Imaging with the full VLBA at 7 mm $\lambda$
shows that the size and shape of Sgr A* is fully consistent
with the predictions of the scattering law derived at 
lower frequencies.  
We find a major axis FWHM of $0.76 \pm 0.04$  mas.
The scattering model predicts $0.67 \pm 0.03$ mas.  We also find an
axial ratio and  a position angle consistent
with that previously identified at longer wavelengths and
with the Backer \etal \markcite{backe93}
(1993) results at 7 mm.  We find no evidence
for structure that is not elliptically symmetric
to a limit of 35 mJy.  
We cannot rule out the existence of a second component in the past:
a synchrotron component at the epoch of the Krichbaum \etal (1993)
observations is likely to have decayed significantly by the epoch of
our observations.

If the intrinsic size adds in quadrature with the scattering size
to form the apparent size, the major and minor axes of the intrinsic source
must be less than
0.48 mas.  
For a galactocentric distance of 8.5 kpc, this
corresponds to 4.1 AU.
We infer a lower limit to
the brightness temperature of $4.9 \times 10^9$ K.
This limit is less than the limit of $1.4 \times 10^{10}$ K
found by Backer \etal \markcite{backe93} (1993) at 7 mm $\lambda$ and
Rogers \etal \markcite{roger94} (1994) at 3 mm $\lambda$ principally
due to the lower flux in our epoch of observation.

The morphology and size of Sgr A* did not change from the
Backer \etal \markcite{backe93} (1993) observations
despite a factor of two decrease in the flux.
The upper limit to source expansion over this epoch is 
$0.04 {\rm\ mas\ y^{-1}} \approx 2 {\rm ~km~s^{-1}}$.
The constant size also
supports the hypothesis that the apparent source is scattered
radiation from the intrinsic source.  An unobscured, homogeneous
synchrotron source with constant magnetic field, spectral
index and peak frequency would show a 40\% increase in its
angular size with a doubling in flux density (Marscher \markcite{marsc83}
1983).
Only a conspiracy of parameter changes could keep the size
from changing significantly.

The constant ellipticity
as a function of wavelength indicates a coherent magnetic field
over 0.02 to $3 \times 10^{-5}$ pc for a scattering screen near
the Galactic Center.  Detection of a change in the ellipticity
with wavelength would be indicative of a turbulent magnetic
field on these scales 
(e.g., Wilkinson, Narayan \& Spencer \markcite{wilki94}
1994).
At 7 mm $\lambda$ the refractive time scale is on the order of 2 years.
Hence, future observers might expect to see a change in the ellipticity
of Sgr A*.

These results are consistent with the hypothesis that the radio
emission from Sgr A* results from synchrotron or
cyclo-synchrotron radiation 
of gas in the vicinity of a black hole.  
Both radial and rotating flow models predict the mass of the black
hole on the order of $10^6 M_{\sun}$ and the accretion rate in the
range $10^{21}$ to $10^{22} {\rm\ g\ s^{-1}}$ (Melia 
\markcite{melia94} 1994, Narayan \etal \markcite{naray95}
1995).  The radio emission may originate in a spherical cloud
or in a low-power inhomogeneous jet.  However, the absence of any
external feature at any wavelength argues against the jet model.

What will shorter wavelength VLBI observations reveal in Sgr A*?
Future global VLBI arrays at 1.3 mm $\lambda$ 
will have an angular resolution of 20 $\mu$as (Wright \& Bower 
\markcite{wrigh97} 1997),
which is on the scale of a few Schwarzschild radii for a 
$10^6 M_{\sun}$ black hole.  At this wavelength scattering will not dominate
since the expected scattering size is 27 $\mu$as.
However, recent measurements of the centimeter to submillimeter spectrum
indicate the presence of two compact sources on two different
size scales (Serabyn \etal \markcite{serab97} 1997,
Falcke \etal \markcite{falck97} 1997).  
If the intrinsic source responsible for the centimeter
to millimeter wavelength spectrum has a spectral turnover near 3 mm $\lambda$,
then this component may be forever invisible to VLBI.  

\acknowledgements The authors thank H. Falcke and K. Kellermann for
useful discussions.

\newpage

\figcaption[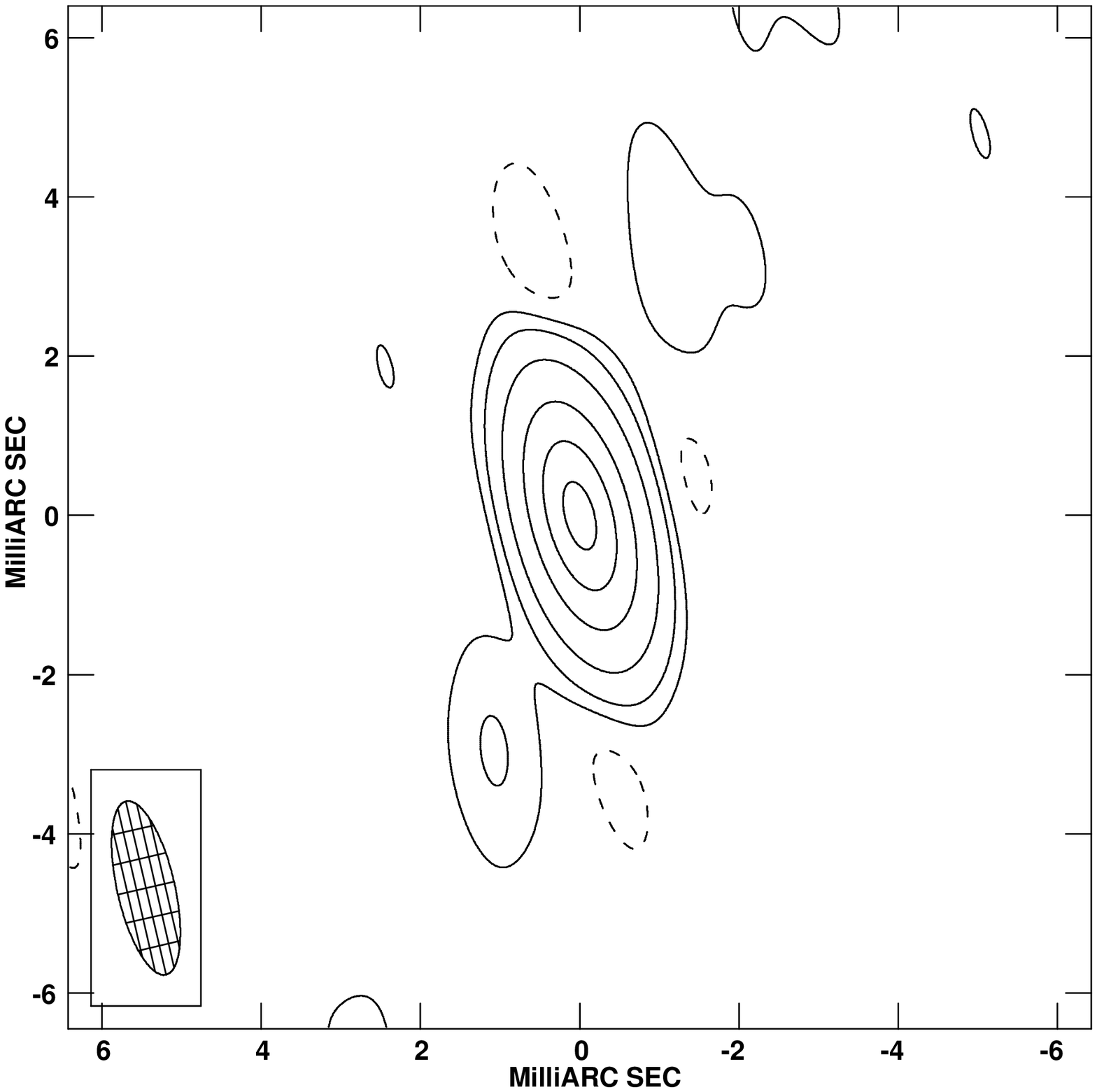]{A uniformly-weighted image of Sgr A*.  The beam is shown
in the lower left hand corner.  The contours are -0.01, 0.01, 
0.03, 0.10, 0.30, 0.60
and 0.90 times the peak intensity of 0.87 Jy/beam.
\label{fig:sgra}}

\figcaption[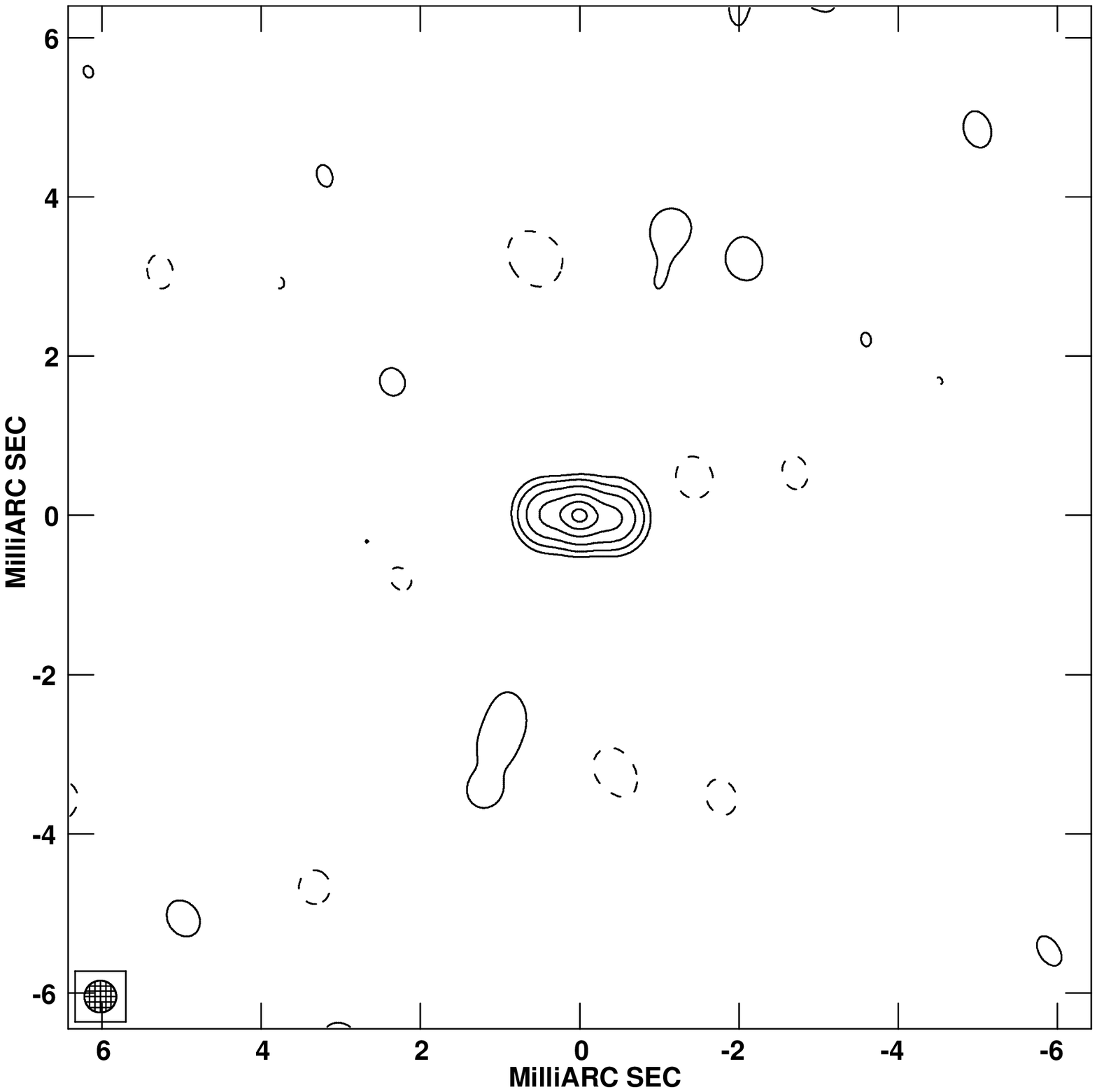]{An image of Sgr A* in which the CLEAN 
components are convolved with
a beam with FWHM of 0.400 mas, shown
in the lower left hand corner.  The contours are -0.01, 0.01,
0.03, 0.10, 0.30, 0.60
and 0.90 times the peak intensity of 0.66 Jy/beam.
\label{fig:sgrasup}}

\figcaption[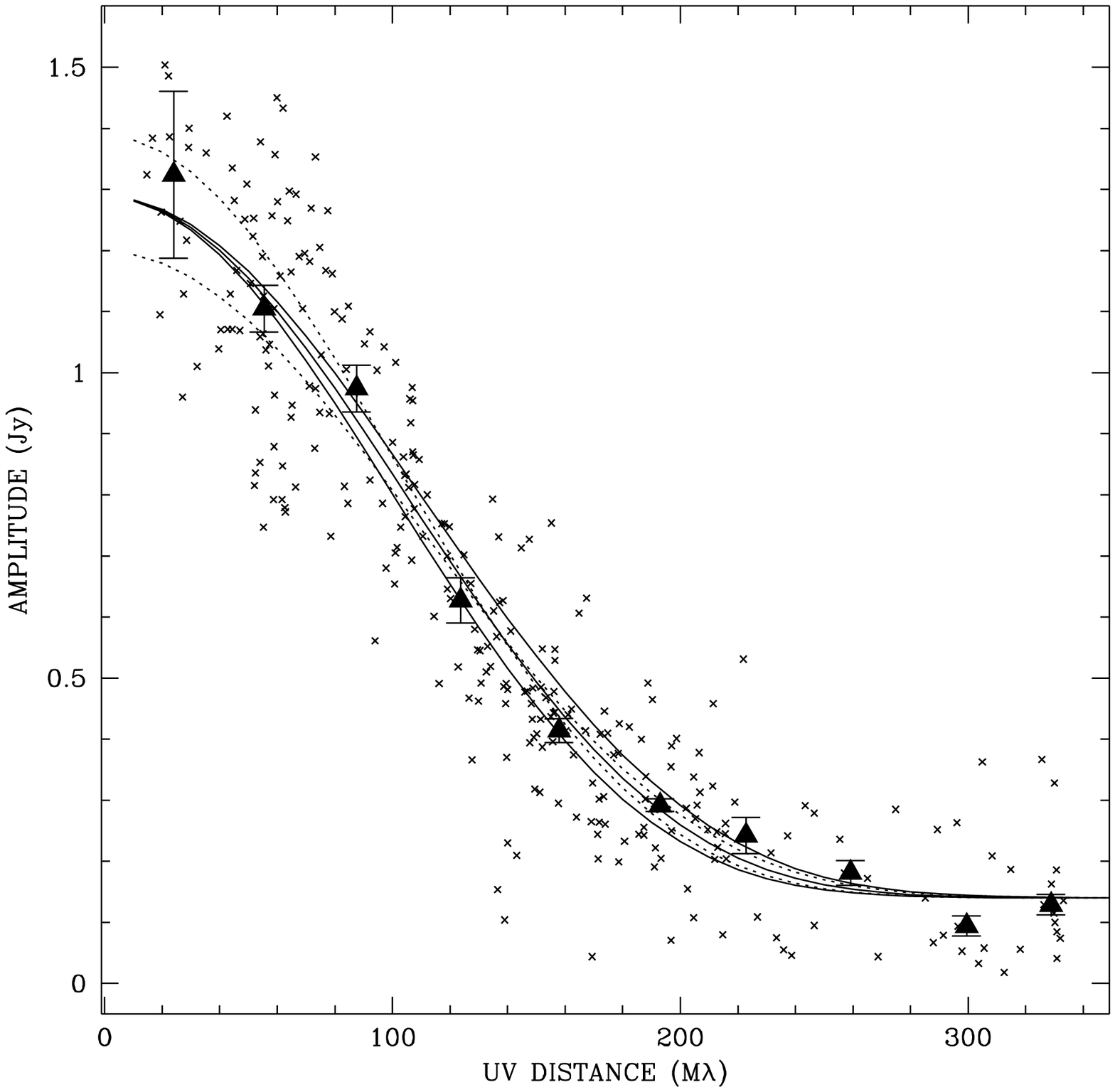]{Visibility amplitude as a function 
of $uv$ distance for 
Sgr A*.  The solid lines indicate the expectation for a circular
Gaussian with a zero baseline flux of 1.28 Jy and 
FWHM of 0.694, 0.727 and 0.760 mas.  The dashed lines indicate the
expectation for a circular Gaussian with zero baseline fluxes of 1.38
Jy and 1.18 Jy with FWHM of 0.760 and 0.694 mas, respectively.
The noise bias is added in quadrature to the model.
The large triangles indicate the median flux in
a 35 M$\lambda$ bin.  The errors 
indicate the scatter in the data.
\label{fig:sgrauvd}}

\figcaption[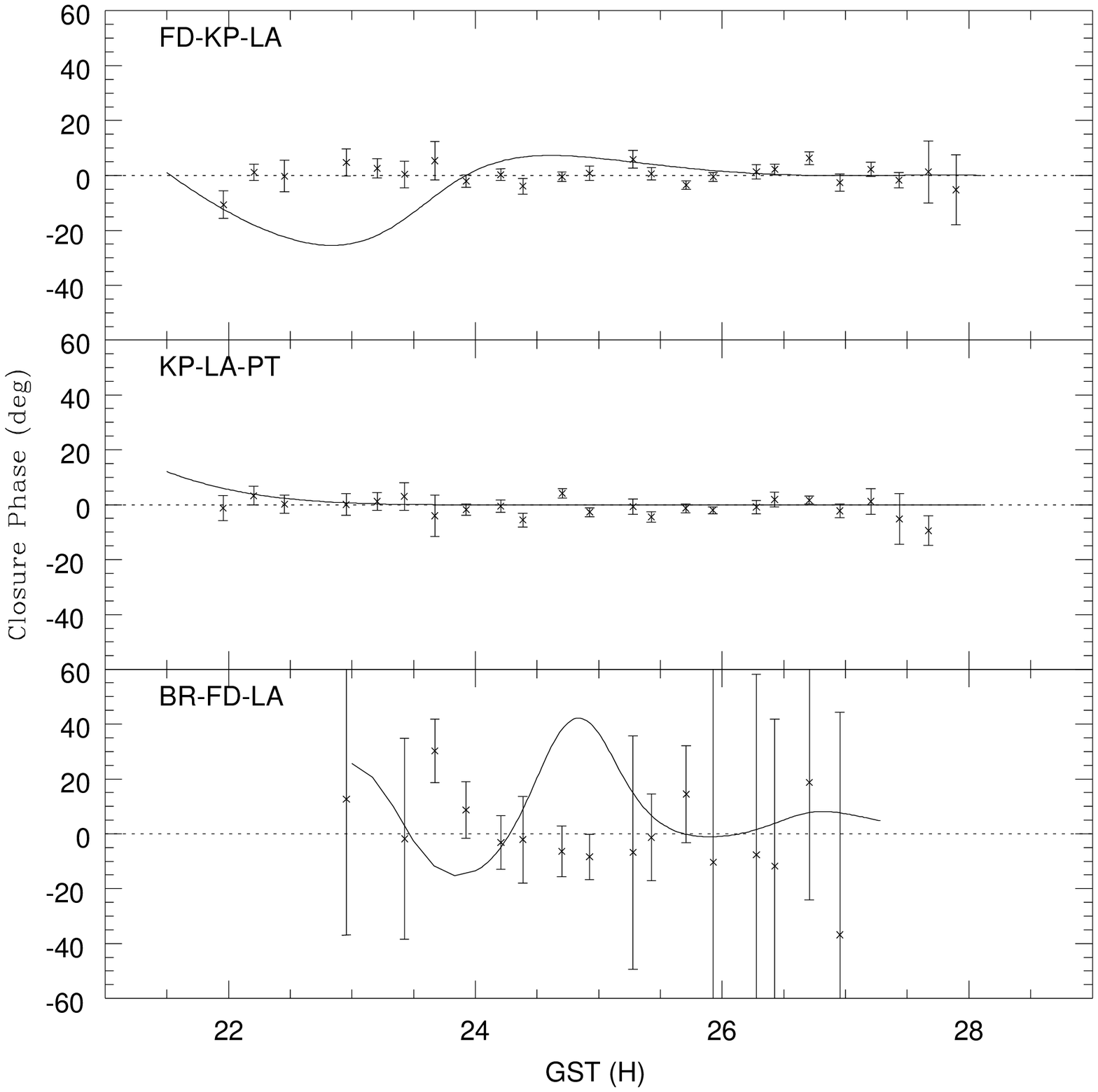]{Closure phase on 3 triangles for Sgr A*.
The solid line is the Krichbaum \etal model.
\label{fig:sgracp1}}

\begin{figure}[p]
\mbox{\psfig{figure=robust.ps,width=14cm}}
\end{figure}

\begin{figure}[p]
\mbox{\psfig{figure=robust_super.ps,width=14cm}}
\end{figure}

\begin{figure}[p]
\mbox{\psfig{figure=best_modbias.ps,width=14cm}}
\end{figure}

\begin{figure}[p]
\mbox{\psfig{figure=paperfig.ps,width=13cm}}
\end{figure}

\end{document}